\documentclass[prd, twocolumn, showpacs, amsmath, amssymb, superscriptaddress, floatfix, nofootinbib]{revtex4}

\usepackage{amsfonts}
\usepackage[dvipdfm, colorlinks=true, citecolor=blue, linkcolor=blue, urlcolor=blue]{hyperref}
\usepackage{color}
\usepackage{graphicx,amsfonts,multirow}
\usepackage{amsmath}

\begin{document}

\title{The $D\to \rho$ semileptonic and radiative decays within the light-cone sum rules}

\author{Hai-Bing Fu}
\email{fuhb@cqu.edu.cn}
\author{Long Zeng}
\email{zenglongz@outlook.com}
\author{Rong L\"{u}}
\email{lvrong97@aliyun.com}
\address{Institute of Particle Physics $\&$ Department of Physics, Guizhou Minzu University, Guiyang 550025, P.R. China}
\author{Wei Cheng}
\email{chengwei@itp.ac.cn}
\address{State Key Laboratory of Theoretical Physics, Institute of Theoretical Physics,
Chinese Academy of Sciences, Beijing, 100190, P.R. China}
\author{Xing-Gang Wu}
\email{wuxg@cqu.edu.cn}
\address{Department of Physics, Chongqing University, Chongqing 401331, P.R. China}

\begin{abstract}
The measured branching ratio of the $D$ meson semileptonic decay $D \to \rho e^+ \nu_e$, which is based on the $0.82~{\rm fb^{-1}}$ CLEO data taken at the peak of $\psi(3770)$ resonance, disagrees with the traditional SVZ sum rules analysis by about three times. In the paper, we show that this discrepancy can be eliminated by applying the QCD light-cone sum rules (LCSR) approach to calculate the $D\to \rho$ transition form factors $A_{1,2}(q^2)$ and $V(q^2)$. After extrapolating the LCSR predictions of these TFFs to whole $q^2$-region, we obtain $1/|V_{\rm cd}|^2 \times \Gamma(D \to \rho e \nu_e) =(55.45^{+13.34}_{-9.41})\times 10^{-15}~{\rm GeV}$. Using the CKM matrix element and the $D^0(D^+)$ lifetime from the Particle Data Group, we obtain ${\cal B} (D^0\to \rho^- e^+ \nu_e) = (1.749^{+0.421}_{-0.297}\pm 0.006)\times 10^{-3}$ and ${\cal B} (D^+ \to \rho^0 e^+ \nu_e) = (2.217^{+0.534}_{-0.376}\pm 0.015)\times 10^{-3}$, which agree with the CLEO measurements within errors. We also calculate the branching ratios of the two $D$ meson radiative processes and obtain ${\cal B}(D^0\to \rho^0 \gamma)= (1.744^{+0.598}_{-0.704})\times 10^{-5}$ and ${\cal B}(D^+ \to \rho^+ \gamma) = (5.034^{+0.939}_{-0.958})\times 10^{-5}$, which also agree with the Belle measurements within errors. Thus we think the LCSR approach is applicable for dealing with the $D$ meson decays.
\end{abstract}


\maketitle

\section{Introduction}

The semileptonic decays of the heavy meson, which contains heavy $c$ or $b$ quark, are important for studying the weak and strong interactions and for studying the heavy-flavor physics. In the charm factories nowadays, such as Belle, LHCb, BES and PANDA, the $D$ meson semileptonic decays provide a good platform for the precision test of standard model (SM) and for searching of new physics (NP) beyond the SM. For examples, the CLEO Collaboration present the first measurement of the branching fraction of $D^+\to \omega e^+\nu_e$~\cite{Coan:2005iu, Huang:2005iv}. Lately, the CLEO Collaboration finished a more precise measurement on the $D^0 \to \rho^- e^+ \nu_e$ and $D^+ \to \rho^0 e^+ \nu_e$ decays based on $0.82~{\rm fb^{-1}}$ data taken at the peak of $\psi(3770)$ resonance. The CLEO Collaboration gave the branching fractions ${\cal B}(D^0 \to \rho^- e^+ \nu_e) = (1.77\pm0.12\pm0.10)\times 10^{-3}$ and ${\cal B}(D^+ \to \rho^0 e^+ \nu_e) = (2.17\pm0.12^{+0.12}_{-0.22})\times 10^{-3}$~\cite{CLEO:2011ab}. Recently, the BES-III Collaboration published their improved results, $\mathcal B (D^0 \to \rho^- e^+ \nu_e) = ( 1.445 \pm 0.058 \pm 0.039) \times 10^{-3}$ and ${\cal B}(D^+ \to \rho^0 e^+ \nu_e) = (1.860 \pm 0.070 \pm 0.061) \times 10^{-3}$, by using a data sample corresponding to an integrated luminosity of 2.93 fb$^{-1}$ at a center-of-mass energy of 3.773 GeV~\cite{Ablikim:2018qzz}.

Meanwhile, the nonleptonic decays of the charm meson such as $D^0\to\rho^0\gamma$ and $D^+\to\rho^+\gamma$  have also been a subject of high priority for many years and has resulted in a wealth of experimental data. In particular, two measured observables by the Belle Collaboration are, ${\cal B}(D^0\to\rho^0\gamma) = (1.77\pm0.30\pm0.07) \times 10^{-5}$ and $A_{CP}(D^0\to\rho^0\gamma) = 0.056\pm0.152\pm0.006$~\cite{Abdesselam:2016yvr}, where the charge-parity (CP) asymmetry $A_{CP} $ is defined as
\begin{equation}
A_{CP}(D\to\rho\gamma) = \frac{\Gamma(D\to \rho\gamma) - \Gamma(\bar D\to \bar \rho\gamma)}{\Gamma(D\to \rho\gamma) + \Gamma(\bar D\to \bar \rho\gamma)} .
\end{equation}

The $D\to\rho$ transition form factors (TFFs) are key components of the $D$ meson semileptonic decay $D\to \rho  e \nu_e $ and the $D$ meson radiative decay $D\to \rho \gamma$. In order to achieve more precise predictions on the $D$ meson semileptonic and radiative decays, it is important to have more accurate $D\to\rho$ TFFs. Theoretically, the $D\to \rho$ TFFs have been calculated under various approaches, such as the 3-point QCD sum rules (3PSR)~\cite{Ball:1993tp}, the heavy quark effective field theory (HQEFT)~\cite{Wang:2002zba, Wu:2006rd}, the relativistic harmonic oscillator potential model (RHOPM)~\cite{Wirbel:1985ji}, the quark model (QM)~\cite{Isgur:1988gb,Melikhov:2000yu}, the light-front quark model (LFQM)~\cite{Verma:2011yw}, the heavy meson and chiral symmetries (HM$\chi$T)~\cite{Fajfer:2005ug}, and the Lattice QCD~\cite{Lubicz:1991bi,Bernard:1991bz}. Most results are consistent with the CLEO measurements within errors; while the QCD sum rules (SR) leads to much smaller branching ratios, i.e. ${\cal B}(D^0\to \rho^- e^+ \nu_e) = (0.5\pm0.1)\times 10^{-3}$~\cite{Ball:1993tp}. One may question the applicability of the QCD SR approach for those TFFs. There are large uncertainties for the 3PSR prediction, which is however based on the conventional Shifman-Vainshtein-Zakharov SR approach~\cite{Shifman:1978bx} and the approach itself has many defaults in dealing with such kind of TFFs~\cite{Braun:1997kw}.

The QCD SR approach is applicable in both small and intermediate $q^2$-region, which provides a bridge between the pQCD and lattice QCD predictions, and then a more accurate QCD SR prediction is helpful for a better understanding of the TFFs. In the paper, as a new QCD SR analysis, we take the improved version of QCD SR approach, i.e. the light-cone sum rules (LCSR)~\cite{Balitsky:1989ry, Chernyak:1990ag, Ball:1991bs}, to recalculate the $D\to\rho$ TFFs. The LCSR avoids the problems of the conventional SVZ SR by making a partial resummation of the operator product expansion (OPE) to all orders and reorganize the OPE expansion in terms of the twists of relevant operators rather than their dimensions. The vacuum condensates of the SVZ SR are then substituted by the light-meson¡¯s light-cone distribution amplitudes (LCDAs) of increasing twists. The LCDA, which relates the matrix elements of the nonlocal light-ray operators sandwiched between the hadronic state and the vacuum, has a direct physical significance and provides the underlying links between the hadronic phenomena at small and large distances. Since its invention, the LCSR approach has been widely adopted for studying the heavy $\to$ light meson decays.

The remaining parts of the paper are organized as follows. In Sec.~\ref{section:2}, we present the calculation technology for the $D\to \rho$ semileptonic and radiative decays, and their key components, e.g. the $D\to \rho$ TFFs, within the LCSR approach. In Sec.~\ref{section:3}, we present our numerical results and discussions on the $D\to \rho$ TFFs, the $D\to \rho$ semileptonic decay widthes and branching ratios for two different channels, and the branching ratio and the direct CP asymmetry of the $D\to\rho\gamma$ decay. Section~\ref{section:4} is reserved for a summary.

\section{Calculation technology}\label{section:2}

\subsection{The $D\to\rho e^+ \nu_e$ semileptonic decay}

\begin{figure}[htb]
  \centering
  \includegraphics[width=0.35\textwidth]{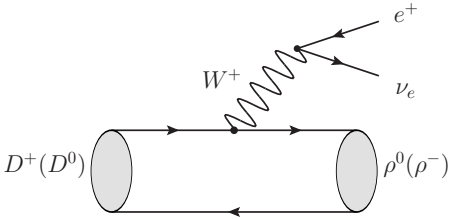}\\
  \caption{Typical diagram for the $D\to\rho e^+ \nu_e$ semileptonic decay.}\label{semidecay}
\end{figure}

We present the typical diagram for the $D\to\rho e^+ \nu_e$ semileptonic decay in Fig.\ref{semidecay}. The longitudinal and transverse helicity decay widths of $D\to \rho e^+ \nu_e$ can be expressed in terms of three helicity amplitudes $H_{0,\pm}(q^2)$:
\begin{eqnarray}
\Gamma^a &=&\frac{G_F^2 |K_{\rm HV}|^2 }{192\pi^3 m_D^3}\int_{m_e^2}^{q^2_{\rm max}}q^2\sqrt{\lambda(q^2)} |H_a(q^2)|^2, \label{Eq:GammaLT0}
\end{eqnarray}
where $a=+$, $-$, $0$, $q^2_{\rm max} = (m_D-m_\rho)^2$, and $\lambda(q^2)$ is the phase-space factor, which equals to $(m_D^2 + m_\rho^2 - q^2)^2-4 m_D^2 m_\rho^2$. The Fermi constant $G_F=1.166\times10^{-5}\;{\rm GeV}^{-2}$. The constant $K_{\rm HV}$ parameterizes the quark flavor mixing relevant to a particular transition, which equals to $|V_{\rm cd}|$ for $D^0 \to \rho^- e^+ \nu_e$ and $|V_{\rm cd}|/{\sqrt 2}$ for $D^+ \to \rho^0 e^+ \nu_e$. Then the total decay width of $D^{+/0}\to \rho^{0/-} e^+ \nu_e$ can be written as
\begin{equation}
\Gamma=\Gamma^{\rm L} + \Gamma^{\rm T},  \label{Eq:Gamma}
\end{equation}
where $\Gamma^{\rm T} = \Gamma^+ + \Gamma^-$ and $\Gamma^{\rm L}=\Gamma^0$.

Here we have adopted the helicity basis to express the decay width. In the helicity basis, the TFF $H_a(q^2)$ corresponds to a transition amplitude with definite spin-parity quantum number in the lepton pair center-of-mass frame. The transverse and longitudinal helicity TFF $H_{a}(q^2)$ can be calculated by relating them to the TFFs $A_{1,2}(q^2)$ and $V(q^2)$ via the following way
\begin{eqnarray}
H_\pm(q^2) &=& (m_D+m_\rho) A_1(q^2)\mp\frac{\sqrt{\lambda(q^2)}} {m_D + m_\rho}V(q^2) \label{Hpm}
\end{eqnarray}
and
\begin{eqnarray}
H_0(q^2) &=& \frac{1}{2m_\rho\sqrt{q^2}}\bigg\{(m_D^2-m_\rho^2-q^2) (m_D+m_\rho)  A_1(q^2)
\nonumber\\
&& -\frac{\lambda(q^2)}{m_D+m_\rho}A_2(q^2)\bigg\}. \label{H0}
\end{eqnarray}
The TFFs $A_{1,2}(q^2)$ and $V(q^2)$ can be defined by relating them to the $D\to\rho$ matrix elements, e.g.
\begin{eqnarray}
&&\langle \rho (\tilde p,\tilde \epsilon )|\bar q \gamma_\mu c|D(p)\rangle = \epsilon_{\mu\nu\alpha\beta}\tilde \epsilon^{*\nu} p^\alpha \tilde p^\beta \frac{2V(q^2)}{m_D + m_\rho },   \label{Drho:matrix1}    \\
&&\langle \rho (\tilde p,\tilde \epsilon )|\bar q{\gamma _\mu } {\gamma _5}c|D(p)\rangle = i \tilde \epsilon^*_\mu (m_D + m_\rho )A_1(q^2)
\nonumber\\
&&\qquad - i(\tilde \epsilon^* \cdot p )(p + \tilde p)_\mu\frac{1}{m_D + m_\rho}A_2(q^2)
\nonumber\\
&&\qquad - i(\tilde \epsilon^* \cdot p )(p - \tilde p)_\mu \frac{2 m_\rho } {q^2}[A_3(q^2) - A_0(q^2)]. \label{Drho:matrix2}
\end{eqnarray}
where $\tilde \epsilon$ is $\rho$-meson polarization vector and $q$ is the four-momentum transfer between the two mesons.

\subsection{The $D\to \rho\gamma$ decay}

We give a mini-review of the Cabibbo-suppressed $D\to\rho\gamma$ decay in this subsection, which is an important channel for testing the SM and for searching of new physics beyond the SM.

The amplitude of $D\to \rho\gamma$ decay can be decomposed into the following gauge invariant form,
\begin{eqnarray}
&&{\cal A}[D(p)\to\rho(\tilde p, \tilde \epsilon)\gamma(q, \epsilon)]=  \epsilon_{\mu\nu\alpha\beta} q^{\mu} \epsilon^{*\nu} p^{\alpha}\tilde\epsilon^{*\beta} {\cal A}_{\rm PC}^\rho
\nonumber\\
&& \qquad + i [(\tilde\epsilon^*\cdot q)(\epsilon\cdot\tilde p)-(\tilde p\cdot q)(\tilde \epsilon^*\cdot \epsilon^*)] {\cal A}_{\rm PV}^\rho ,
\end{eqnarray}
where $\tilde\epsilon$ and $\epsilon$ are polarization vectors of $\rho$ and $\gamma$, respectively. ${\cal A}_{\rm PC}^\rho $ and ${\cal A}_{\rm PV}^\rho $ are parity conserving and parity violating amplitudes which can be calculated by using the effective $c\to u\gamma$ weak Lagrangian~\cite{Burdman:1995te, Greub:1996wn, Fajfer:1997bh, Isidori:2012yx}. Then the decay rate of $D\to \rho\gamma$ can be written as~\cite{deBoer:2015boa, deBoer:2017que}
\begin{eqnarray}
\Gamma(D\to \rho\gamma) = \frac{m_D^3}{32 \pi} \bigg(1-\frac{m_\rho^2}{m_D^2}\bigg)^3 \big[|{\cal A}_{\rm PV}^\rho |^2 + |{\cal A}_{\rm PC}^\rho |^2 \big].
\end{eqnarray}

Our remaining task is to calculate the amplitudes ${\cal A}_{\rm PV, PC}^\rho$, which contain both the long-distance contribution and the short-distance contribution, e.g.
\begin{equation}
{\cal A}_{\rm PV, PC}^\rho=({\cal A}_{\rm PV, PC}^\rho)^{\rm l.d.}+ ({\cal A}_{\rm PV, PC}^\rho)^{\rm s.d.}. \label{Eq:APCV_ld}
\end{equation}

\begin{figure}[htb]
  \centering
  \includegraphics[width=0.35\textwidth]{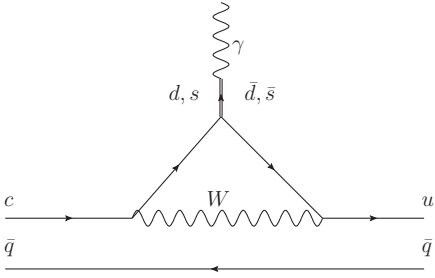}\\
  \caption{The typical VME diagram via the penguin-like $c\to u\gamma$ transition.}\label{Fig:Fig1}
\end{figure}

\begin{figure*}
  \includegraphics[width=0.9\textwidth]{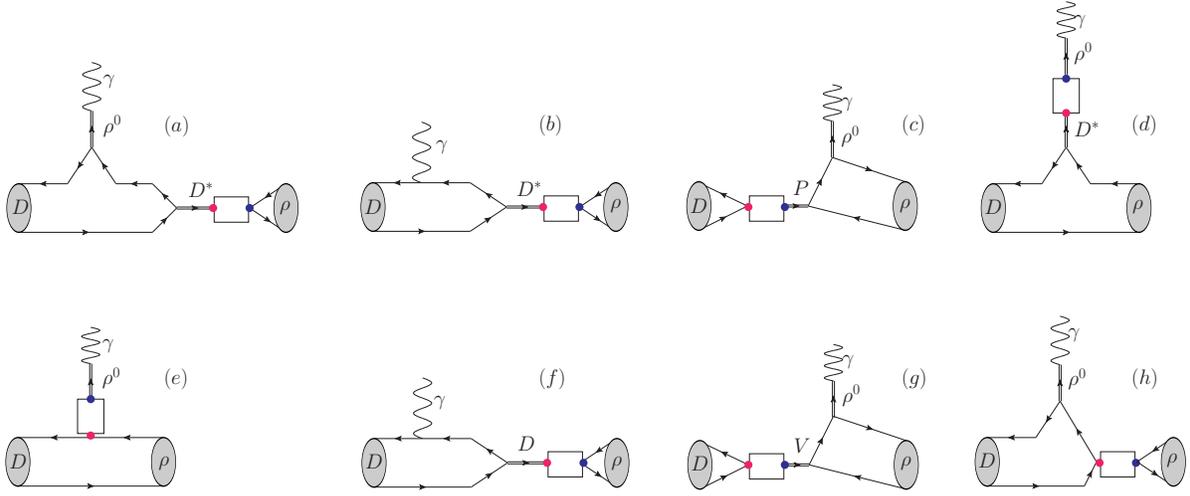}\\
  \caption{Skeleton diagrams for the long-distance amplitudes which contribute to the $D\to\rho\gamma$ decay. The rectangle denotes the weak transition which can be described with the help of the effective Lagrangian. }\label{Fig:Fig2}
\end{figure*}

Firstly, the long-distance contributions for radiative decays $D\to\rho\gamma$ can be treated as originating from the nonleptonic transition $D\to \rho V_0$, followed by the conversion $V_0\to \gamma$ via the vector meson dominance (VME) mechanism~\cite{Fajfer:1998dv}, whose typical Feynman diagram via the penguin-like $c\to u\gamma$ transition is shown in Fig.\ref{Fig:Fig1}. More precisely, the long-distance amplitudes $({\cal A}_{\rm PV, PC}^\rho)^{\rm l.d.}$ can be divided into three parts, e.g.
\begin{displaymath}
({\cal A}_{\rm PV, PC}^\rho)^{\rm l.d.} = {\cal A}_{\rm PV, PC}^{\rm I} +{\cal A}_{\rm PV, PC}^{\rm II} +{\cal A}_{\rm PV, PC}^{\rm III}
\end{displaymath}
with $8$ schematic diagrams to be calculated, are the hybrid model based on the heavy quark effective theory and chiral perturbation theory. The rectangle in each diagram of Fig.\ref{Fig:Fig2} respects the weak transition due to the effective Lagrangian which can be found in Ref.\cite{Fajfer:1998dv}. The Lagrangian contains a product of two left-handed quark currents, which will be expressed in terms of the relevant hadronic degrees of freedom, $D$, $D^*$, $P$ and $V$, as described in Fig.\ref{Fig:Fig2}:
\begin{itemize}
\item The first part ${\cal A}_{\rm PC}^{\rm I}$, which is represented by Fig.\ref{Fig:Fig2}(a) and Fig.\ref{Fig:Fig2}(b), denotes the photon emitted from the $D$ meson becomes $D^*$ meson and then weakly decays into $\rho$ meson.
\item The second part ${\cal A}_{\rm PC}^{\rm II}$, which is shown by Fig.\ref{Fig:Fig2}(c), comes from the weak decay of $D$ meson, which firstly decays into an off-shell light pseudoscalar and then decays into $\rho\gamma$.
\item As shown by Fig.\ref{Fig:Fig2}(d) and Fig.\ref{Fig:Fig2}(e), there are also long-distance penguin-like contributions, which contribute to both the parity-conserving amplitude ${\cal A}_{\rm PC}^{\rm III}$ and the parity-violating amplitude ${\cal A}_{\rm PV}^{\rm III}$. They will be vanished in the exact SU(3) flavour limit.
\item Fig.\ref{Fig:Fig2}(f) and Fig.\ref{Fig:Fig2}(g) are bremsstrahlung-like diagrams, where the photon emission is due to the direct coupling to charged initial $D$ or final-state $\rho$-meson, both of which contribute to the parity violating amplitude ${\cal A}_{\rm PV}^{\rm I}$.
\item Fig.\ref{Fig:Fig2}(h) denotes the remaining parity violating contribution from other quark-level picture~\cite{Fajfer:1998dv}, whose amplitude is represented as ${\cal A}_{\rm PV}^{\rm II}$.
\end{itemize}
It has been found that Fig.\ref{Fig:Fig2}(c) has sizable contribution, while Figs.\ref{Fig:Fig2}(d) and (e) have negligible contributions which are less than $1\%$. To derive the amplitudes ${\cal A}_{\rm PV, PC}^{\rm I, II, III}$, we need to know the coupling strengthes for the interaction of the light vector meson with $D$ or $D^*$ and the heavy quark-photon interaction, which can be characterized by the parameters $\lambda$ and $\lambda'$, respectively. To do our numerical calculation, we shall adopt $\lambda= 0.479 ~{\rm GeV}^{-1}$~\cite{Fajfer:1998dv}, and $\lambda' = 1/(6m_c)$ which is predicted by approximately relating it to the charm quark magnetic moment~\cite{Isgur:1989vq, Yan:1992gz, Cho:1992nt, Amundson:1992yp}.

Following the above discussions, the parity-conserving amplitudes for $D^+ \to \rho^+ \gamma$ are
\begin{eqnarray}
{\cal A}_{\rm PC}^{\rm I} (D^+ \to \rho^+ \gamma) &=& 4a_1 {\cal G} \bigg| \lambda' - \lambda \frac{{\tilde g}_V}{2{\sqrt 2}} \left(\frac{f_\rho}{m_\rho}
- \frac{f_\omega}{3 m_\omega}\right) \bigg|
\nonumber\\
&&\times \frac{m_\rho m_{D^*} f_D f_\rho}{m_{D^*}^2 - m_\rho^2}
{\sqrt \frac{m_{D^*}}{m_D}},
\nonumber\\
\nonumber\\
{\cal A}_{\rm PC}^{\rm II} (D^+ \to \rho^+ \gamma) &=& 4a_1 {\cal G}
\frac{m_{D}^2 f_D f_\omega }{3 m_\omega (m_{D}^2 - m_{\pi}^2)} |C_{VV\Pi}|,
\nonumber\\
\nonumber\\
{\cal A}_{\rm PC}^{\rm III} (D^+ \to \rho^+ \gamma) &=& a_2 {\cal G}
\left(-f_\rho^2 + \frac1{3}f_\omega^2 - \frac{V_{cs}^*V_{us}}{V_{cd}^*V_{ud}}\frac{2}{3}f_\phi^2\right)
\nonumber\\
&&\times \frac{|V(0)|}{m_D + m_\rho}
\label{apcdpcs}
\end{eqnarray}
and the parity-violating amplitudes for $D^+ \to \rho^+ \gamma$ are
\begin{eqnarray}
&& {\cal A}_{\rm PV}^{\rm I} (D^+ \to \rho^+ \gamma) = 2a_1{\cal G}\frac{m_\rho f_{\rho} f_D  }{m_D^2 - m_{\rho}^2},
\nonumber\\
\nonumber\\
&& {\cal A}_{\rm PV}^{\rm II} (D^+ \to \rho^+ \gamma) = - a_1{\cal G} \frac{f_\rho}{m_D^2 - m_\rho^2} \bigg[f_\rho (m_D + m_\rho)
\nonumber\\
&&\qquad\qquad  \times A_1(m_\rho^2)  - f_\omega \frac{m_\rho(m_D + m_\omega)}{3m_\omega} A_1^{D\omega}(m_\rho^2)\bigg],
\nonumber\\
\nonumber\\
&& {\cal A}_{\rm PV}^{\rm III} (D^+ \to \rho^+ \gamma) = a_2 {\cal G} \left(-f_\rho^2 + \frac1{3}f_\omega^2 - \frac{V_{cs}^*V_{us}}{V_{cd}^*V_{ud}}\frac{2}{3}f_\phi^2\right)
\nonumber\\
&&\qquad\qquad  \times \frac{A_1(0)}{m_D - m_\rho},
\label{apvdpcs}
\end{eqnarray}
where ${\cal G} = e G_F V_{ud} V_{cd}^*/\sqrt 2$ with $e = \sqrt{4\pi \alpha_{\rm em}}$, $\tilde{g}_{V} = 5.9$ which can be fixed by the flavor symmetry, $|C_{VV\Pi}| = 3 \tilde{g}_{V}^2/(32\pi^2) = 0.33$~\cite{Fajfer:1997bh}, $a_1 = 1.26\pm0.1$ and $a_2 = -0.55\pm0.1$~\cite{Bauer:1986bm}, $f_{\rho,\omega,\phi}$ are decay constants which can be fixed by the leptonic decays of these mesons. $A_1^{D\omega}(m_\rho^2)$ is the TFF for $D\to\omega$~\cite{Fajfer:1997bh}.

The Cabibbo suppressed decay $D^0 \to \rho^0 \gamma$ involves the contribution from the $\eta - \eta^{\prime}$ mixing, whose parity-conserving amplitudes are
\begin{eqnarray}
{\cal A}_{\rm PC}^{\rm I} (D^0 \to \rho^0 \gamma) &=&  4 a_2 b_\rho^0 {\cal G}
\bigg| \lambda^{\prime} +\lambda \frac{{\tilde g}_V}{2{\sqrt 2}} \left(\frac{f_\rho}{m_\rho} + \frac{f_{\omega}}{3 m_{\omega}}\right) \bigg|
\nonumber\\
&&\times \frac{m_\rho m_{D^*} f_D f_\rho }{m_{D^*}^2 - m_\rho^2} {\sqrt \frac{m_{D^*}}{m_D}},
\nonumber\\
\nonumber\\
{\cal A}_{\rm PC}^{\rm II} (D^0 \to \rho^0 \gamma) &=&  4a_2{\cal G} {\cal B}^\rho m_{D}^2   f_D | C_{VV\Pi} |,
\nonumber\\
\nonumber\\
{\cal A}_{\rm PC}^{\rm III} (D^0 \to \rho^0 \gamma) &=& a_2{\cal G} \left( -f_\rho^2 + \frac1{3}f_\omega^2 - \frac{V_{cs}^*V_{us}}{V_{cd}^*V_{ud}}\frac{2}{3}f_\phi^2 \right)
\nonumber\\
&&\times\frac{V(0)}{m_D + m_\rho},
\label{apcd0cs}
\end{eqnarray}
where $b_{\rho}^0 = - 1/{\sqrt 2}$ and the coefficients
\begin{equation}
{\cal B}^\rho= \sum_{i=1}^{3} \frac{B^{\rho}_{P_i}}{m_D^2 - m_{P_i}^2}, \label{bv}
\end{equation}
where $P_i$ stands for the light mesons $\pi$, $\eta$, and $\eta'$, respectively, and $B^\rho_\pi = f_\omega/(3 \sqrt 2 m_\omega)$, $B^{\rho}_\eta = -c(c - \sqrt 2 s)f_\rho /(\sqrt 2 m_\rho)$, $B^{\rho}_{\eta'} =  -  s ({\sqrt 2} c + s) f_{\rho} /(\sqrt 2 m_{\rho})$ with $s=\sin \theta$ and $c=\cos\theta$. Here $\theta$ is the $\eta-\eta'$ mixing angle and we set its value as $\theta = -20^{\rm o}$, which is consistent with the values derived in Refs.\cite{Tanabashi:2018oca, Huang:2006as, Klempt:2007cp, Aaij:2014jna}. The parity-violating amplitudes for $D^0 \to \rho^0 \gamma$ are
\begin{eqnarray}
&& {\cal A}_{\rm PV}^{\rm I} (D^0 \to \rho^0 \gamma) = 0,
\nonumber\\
\nonumber\\
&& {\cal A}_{\rm PV}^{\rm II} (D^0 \to \rho^0 \gamma)= - a_2{\cal G} \frac{f_\rho}{\sqrt 2( m_D^2 - m_\rho^2)} \bigg[f_\rho (m_D + m_\rho)
\nonumber\\
&&\qquad\qquad  \times A_1(m_\rho^2) + f_\omega \frac{m_\rho(m_D + m_\omega)}{3m_\omega} A_1^{D\omega}(m_\rho^2)\bigg],
\nonumber\\
\nonumber\\
&& {\cal A}_{\rm PV}^{\rm III} (D^0 \to \rho^0 \gamma)=  a_2 {\cal G} \left(-f_\rho^2 + \frac1{3}f_\omega^2 - \frac{V_{cs}^*V_{us}}{V_{cd}^*V_{ud}}\frac{2}{3}f_\phi^2\right)
\nonumber\\
&&\qquad \qquad \times \frac{A_1(0)}{m_D - m_\rho}.
\label{apvdp0s}
\end{eqnarray}

Secondly, we need to deal with the short-distance contribution $({\cal A}_{\rm PC(PV)}^\rho)^{\rm s.d.}$. There are ten operators which enter into the weak interactions of the $D\to\rho$ decay~\cite{Burdman:1995te, Greub:1996wn, Fajfer:1997bh, Isidori:2012yx}. It has been observed that due to the GIM suppression and also the small magnitudes of the Wilson coefficients, only the operators ${\cal Q}_7$ and ${\cal Q}'_7$ have sizable contributions to the $D\to\rho$ decay; Thus the short-distance amplitudes $({\cal A}_{\rm PC(PV)}^\rho)^{\rm s.d.}$ have the following form~\cite{Isidori:2012yx, deBoer:2017que}
\begin{eqnarray}
({\cal A}_{\rm PC(PV)}^\rho)^{\rm s.d.} = \frac{\sqrt{4\pi\alpha_e} Q_u G_F m_c}{2\sqrt 2\pi^2} (A_7 + A'_7)T(0) ,
\end{eqnarray}
in which $T(0) = T_{1}(0)= T_{2}(0)$ and the tensor TFFs $T_{1,2}(q^2)$ are defined by
\begin{eqnarray}
&&\langle \rho(\tilde p,\tilde \epsilon)|\bar u q^\nu \sigma_{\mu\nu}(1+\gamma_5)c|D(p)\rangle
\nonumber\\
\nonumber\\
&&\quad = -2i\epsilon_{\mu\nu\alpha\beta} \tilde \epsilon^{*\nu} p^\alpha \tilde p^\beta T_1(q^2)
\nonumber\\
\nonumber\\
&&\quad + \left[(m_D^2-m_\rho^2)\tilde\epsilon^*_\mu - (\tilde\epsilon^* \cdot p) (p + \tilde p)_\mu\right] T_2(q^2)
 \nonumber\\
&&\quad + (\tilde \epsilon^* \cdot p )\left[(p - \tilde p)_\mu - (p + \tilde p)_\mu \frac{q^2}{m_D^2 - m_\rho^2}\right] T_3(q^2). \nonumber\\
\end{eqnarray}
Those two tensor TFFs $T_{1,2}(q^2)$ can be calculated under the LCSR approach. For the needed $T_1$, by using the heavy quark effective field theory, it can be related to the TFFs $V$ and $A_1$ via the following way~\cite{Wu:2006rd}
\begin{displaymath}
T_1 (q^2) = \frac{m_D^2 - m_\rho^2 + q^2}{2m_D(m_D + m_\rho)}V(q^2) + \frac{m_D + m_\rho}{2m_D}A_1(q^2).
\end{displaymath}
Furthermore, the coefficient $A_7^{(\prime)} = C_7^{(\prime){\rm eff}}+\cdots$, where the ellipse stands for the additional contribution from within and outside the SM.

As a final remark, for the Cabibbo suppressed decay $D^0 \to \rho^0 \gamma$, in the limit where the strong phases of the amplitudes have a mild energy dependence, and assuming that we can neglect the weak phase of the long-distance amplitudes, and the $CP$ asymmetry $|A_{CP}|$ is primarily sensitive to direct $CP$ asymmetry $|a_{\rho\gamma}^{\rm dir}|$, i.e.
\begin{eqnarray}
|A_{CP}| \approx |a_{\rho\gamma}^{\rm dir}| = 2\frac{|{\rm Im}({\cal A}_{\rm PC, PV}^\rho)^{\rm s.d.}|}{|({\cal A}_{\rm PC, PV}^\rho)^{\rm l.d.}|}\times |\sin(\Delta\phi_{\rm strong})|. \nonumber \\
\label{Eq:CP_asymmetry}
\end{eqnarray}

\subsection{The $D\to \rho$ TFFs}

\begin{table}[htb]
\centering
\caption{The $\rho$-meson LCDAs with different twist-structures up to $\delta^3$~\cite{Ball:2004rg}, where $\delta \simeq m_\rho/m_c$.} \label{DA_delta}
\begin{tabular}{cccc}
\hline
 ~~~& ~~~twist-2 ~~~ &~~~twist-3 ~~~& ~~~twist-4 ~~~ \\
\hline
~~$\delta^0$~~      & $\phi_{2;\rho}^\bot$  &  &  \\
$\delta^1$         & $\phi_{2;\rho}^\|$ & $\phi_{3;\rho}^\bot, \psi_{3;\rho}^\bot, \Phi_{3;\rho}^\|,\tilde\Phi_{3;\rho }^\bot$  &   \\
$\delta^2$         &  & $\phi_{3;\rho}^\|, \psi_{3;\rho}^\|, \Phi_{3;\rho}^\bot$ & $  \phi_{4;\rho}^\bot ,\psi_{4;\rho}^\bot,\Psi_{4;\rho}^\bot, \widetilde{\Psi} _{4;\rho}^\bot$\\
$\delta^3$         &  &  & $\phi_{4;\rho}^\|,\psi_{4;\rho}^\|$\\
\hline
\end{tabular}
\end{table}

As mentioned in the Introduction, we shall adopt the LCSR approach to calculate the $D\to\rho$ TFFs. The LCSR approach is based on the operator production expansion near the light cone, and in different to the traditional QCD SR approach which parameterizes all the non-perturbative dynamics into vacuum condensates, the LCSR approach parameterizes those non-perturbative dynamics into LCDAs with increasing twists. Due to the complex structures of the $\rho$-meson LCDAs, it is convenient to arrange them by the parameter $\delta \simeq m_\rho/m_c\sim 52\%$~\cite{Ball:2004rg, Ball:1998sk}. A collection of the $\rho$-meson twist-2, twist-3 and twist-4 LCDAs up to $\delta^3$-order are shown in Table~\ref{DA_delta}.

Up to twist-4 level, there are totally fifteen $\rho$-meson LCDAs, all of which, especially the high-twist DAs, are far from affirmation. As a tricky point of the LCSR approach, one may choose proper current for the correlation function (correlator) so as to suppress less certain high-twist terms and improve the accuracy of the LCSR prediction~\cite{Huang:1998gp, Huang:2001xb, Wan:2002hz, Zuo:2006dk, Wu:2007vi, Wu:2009kq}. For example, one can adopt a right-handed current to do the LCSR calculation, i.e., starting from the following chiral correlator
\begin{eqnarray}
\Pi_\mu(\tilde p,q) &=& i\int d^4x e^{iq\cdot x}   \nonumber\\
&\times& \langle\rho (\tilde p,\tilde\epsilon)|{\rm T} \big\{\bar q_1(x)\gamma_\mu(1-\gamma_5)c(x),  j_D^\dag (0)\big\} |0\rangle, \nonumber\\ \label{correlator}
\end{eqnarray}
where the current $j_D^\dag (x)=i \bar c(x)(1 + \gamma_5)q_2(x)$. This chiral correlator highlights the contributions from the chiral-odd LCDAs $\phi_{2;\rho}^\bot$ at the $\delta^0$-order, $\phi_{3;\rho}^\|$, $\psi_{3;\rho}^\|$, $\Phi_{3;\rho}^\bot$, $\phi_{4;\rho}^\bot$, $\psi_{4;\rho}^\bot$, $\Psi_{4;\rho}^\bot$, $\widetilde{\Psi} _{4;\rho}^\bot$ at the $\delta^2$-order; while all the contributions from the chiral-even $\rho$-meson LCDAs are negligible.

Following the standard LCSR procedures, we can obtain the LCSRs for the $D \to \rho$ TFFs $A_{1,2}(q^2)$ and $V(q^2)$ by using the above chiral correlator, which are similar to the $B\to \rho$ TFFs. The $B\to \rho$ TFFs have calculated by various groups under the LCSR approach, and a recent work is done by applying the vacuum-to-$B$-meson correlation function with an interpolating current for the light vector meson~\cite{Gao:2019lta}. The lengthy analytic expressions for the $D \to \rho$ TFFs with the help of the present choice of a chiral correlator can be obtained from Ref.\cite{Fu:2014pba} by replacing the $B$-meson inputs there as the present $D$ meson ones. To short the length of the paper, we shall not list those formulas here and the interesting reader may turn to Ref.\cite{Fu:2014pba} for detail.

Numerically, we observe that the leading-twist terms are dominant for the chiral LCSRs of the TFFs, agreeing well with the usual $\delta$-power counting rule. Thus, those TFFs shall provide us a useful platform for testing the properties of the leading-twist $\phi_{2;\rho}^\bot$ via comparisons with the data or predictions from other theoretical approaches, such as those of Refs.\cite{Ball:2007zt, Choi:2007yu, Forshaw:2012im, Xu:2018mpf}.

More over, it is convenient to define two ratios over the three TFFs $A_{1,2}(q^2)$ and $V(q^2)$,
\begin{eqnarray}
r_V = \frac{V(0)}{A_1(0)} \qquad {\rm and}\qquad r_2=\frac{A_2(0)}{A_1(0)}.
\end{eqnarray}
Those two ratios could suppress the theoretical errors for each TFF within the LCSR approach, and also suppress the differences of the predictions on the TFFs from various approaches. Thus a better comparison with the data can be achieved.

\section{Numerical results and discussion}\label{section:3}
\subsection{Distribiton amplitudes and TFFs}

To do the numerical calculation, we take the decay constant $f_\rho^\bot = 0.165(9) {\rm GeV}$~\cite{Ball:2007zt}. The $\rho$ and $D$ meson masses are taken as $m_\rho=0.775~{\rm GeV}$ and $m_D = 1.864$ GeV~\cite{Tanabashi:2018oca}. The Cabibbo-Kobayashi-Maskawa matrix element $|V_{\rm cd}| = 0.2252 \pm 0.0007$~\cite{Tanabashi:2018oca}, and the $D$ meson decay constant $f_D$ shall be determined by using the QCD sum rules approach~\cite{Fu:2013wqa}. We adopt the WH model~\cite{Wu:2010zc} as the $\rho$-meson transverse leading twist wavefunction, whose radial part is from the BHL-prescription~\cite{BHL} and the spin-space wavefunction $\chi_\rho^{h_1 h_2} (x,{\bf k}_\bot)$ is from Wigner-Melosh rotation. And then, after integrating out the transverse moment dependence, we obtain the $\rho$-meson LCDA
\begin{eqnarray}
&& \phi _{2;\rho }^\bot (x,\mu)  = \frac{{A_{2;\rho}^\bot \sqrt {3x\bar x} {m_q}}}{{8{\pi ^{\frac{3}{2}}}\widetilde f_\rho^\bot b_{2;\rho}^\bot }}[1 + {B_{2;\rho}^\bot }C_2^{\frac{3}{2}}(\xi )]\nonumber\\
&& \quad \times \left[ {{\rm{Erf}}\left( {b_{2;\rho}^\bot \sqrt {\frac{{{\mu^2} + m_q^2}}{{x\bar x}}} } \right) - {\rm{Erf}}\left( {b_{2;\rho}^\bot \sqrt {\frac{{m_q^2}}{{x\bar x}}} } \right)} \right], \nonumber\\  \label{Eq:DAWH}
\end{eqnarray}
where $\widetilde{f}_\rho^\bot = f_\rho^{\bot}/\sqrt{3}$ and the error function, $\textrm{Erf}(x) = 2 \int^x_0 e^{-t^2} dt/ \sqrt \pi$. The constitute light-quark mass is taken as $m_q \simeq 300~{\rm MeV}$. The two parameters $A_{2;\rho}^\bot$ and $b_{2;\rho}^\bot$ can be fixed by the normalization condition and the average value of the squared transverse momentum $\langle {\bf k}_\bot^2 \rangle_{2;\rho}^{1/2} = 0.37 \pm 0.02 ~{\rm GeV}$~\cite{Fu:2014pba}. The parameter $B_{2;\rho}^\bot$ can be fixed by using the second Gegenbauer moment, i.e. $a_{2;\rho}^\bot(\mu_0 = 1~{\rm GeV}) = 0.14(6)$~\cite{Ball:2007zt}.

Using those input values, we obtain
\begin{eqnarray}
&& A_{2;\rho}^{\bot,{\rm C}} = 23.808,~b_{2;\rho}^{\bot,{\rm C}}=0.572,~B_{2;\rho}^{\bot,{\rm C}} = 0.100;  \\
&& A_{2;\rho}^{\bot,{\rm U}} = 22.679,~b_{2;\rho}^{\bot,{\rm U}}=0.555,~B_{2;\rho}^{\bot,{\rm U}} = 0.151;  \\
&& A_{2;\rho}^{\bot,{\rm D}} = 25.212,~b_{2;\rho}^{\bot,{\rm D}}=0.595,~B_{2;\rho}^{\bot,{\rm D}} = 0.050,
\end{eqnarray}
where $C$, $U$ and $D$ stand for center, upper and lower values, respectively.

\begin{figure}[htb]
\centering
\includegraphics[width=0.45\textwidth]{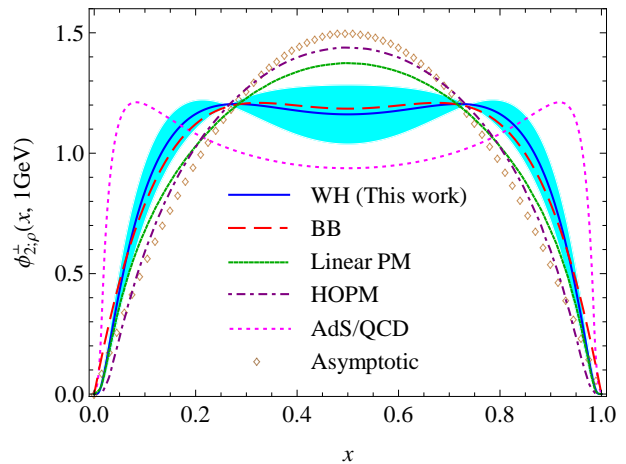}
\caption{The $\rho$-meson leading-twist LCDA $\phi_{2;\rho}^\bot(x,\mu_0 = 1~{\rm GeV})$ predicted from the WH model. As a comparison, the BB prediction~\cite{Ball:2007zt}, the Linear PM and HOPM~\cite{Choi:2007yu}, the AdS/QCD prediction~\cite{Forshaw:2012im}, and the asymptotic form have also been presented.
} \label{Fig:DA}
\end{figure}

We present the $\rho$ meson transverse twist-2 LCDA $\phi_{2;\rho}^\bot(x,\mu_0 = 1~{\rm GeV})$ in Fig.~\ref{Fig:DA}. As a comparison, the Ball and Braun (BB) model~\cite{Ball:2007zt}, the Linear potential model (PM) and the harmonic oscillator potential model (HOPM)~\cite{Choi:2007yu}, the AdS/QCD model~\cite{Forshaw:2012im}, and the asymptotic model have also been presented. Fig.~\ref{Fig:DA} shows that the shape of $\phi_{2;\rho}^\bot$ is still unfixed, which varies from a single peaked behavior to a double peaked behavior. The $\phi_{2;\rho}^\bot$ shape is primarily controlled by the magnitude of $B_{2;\rho}^\bot$ or equivalently the second Gegenbauer moment $a_{2;\rho}^\bot$. Thus the WH model provides a convenient way to fix the behavior of $\phi_{2;\rho}^\bot$ via comparing with the data.

To set the Borel windows for the LCSRs of the $D \to \rho$ TFFs, we adopt the following criteria
\begin{itemize}
\item[(i)] We require the continuum contribution to be less than $30\%$ of the total LCSR.
\item[(ii)] We require all the high-twist LCDAs' contributions to be less than $15\%$ of the total LCSR.
\item[(iii)] The derivatives of LCSRs for TFFs with respect to $(-1/M^2)$ give three LCSRs for the $D$ meson mass $m_{D}$. We require the predicted $D$ meson mass to be fulfilled in comparing with the experiment one, e.g. $|m^{\rm th}_{D}-m^{\rm exp}_{D}|/m^{\rm exp}_{D}$ less than $0.1\%$.
\end{itemize}

We take the continuum thresholds for $D\to\rho$ TFFs $A_{1,2}(q^2)$ and $V(q^2)$ as $s_0(A_1) = 6.1(3)~{\rm GeV}^2$, $s_0(A_2) = 7.1(3)~{\rm GeV}^2$ and $ s_0(V) = 6.6(3)~{\rm GeV}^2$, which are close to the squared mass of the $D$ meson's first excited state $D_1(2420)$. Numerically, we observe that the TFFs change slightly with $s_0$, thus the uncertainties caused by different choices $s_0$ are small~\footnote{Such a small $s_0$ dependence also plays a role to suppress the unwanted scalar contribution due to the choice of chiral correlator. }.

\begin{figure}[htb]
\centering
\includegraphics[width=0.45\textwidth]{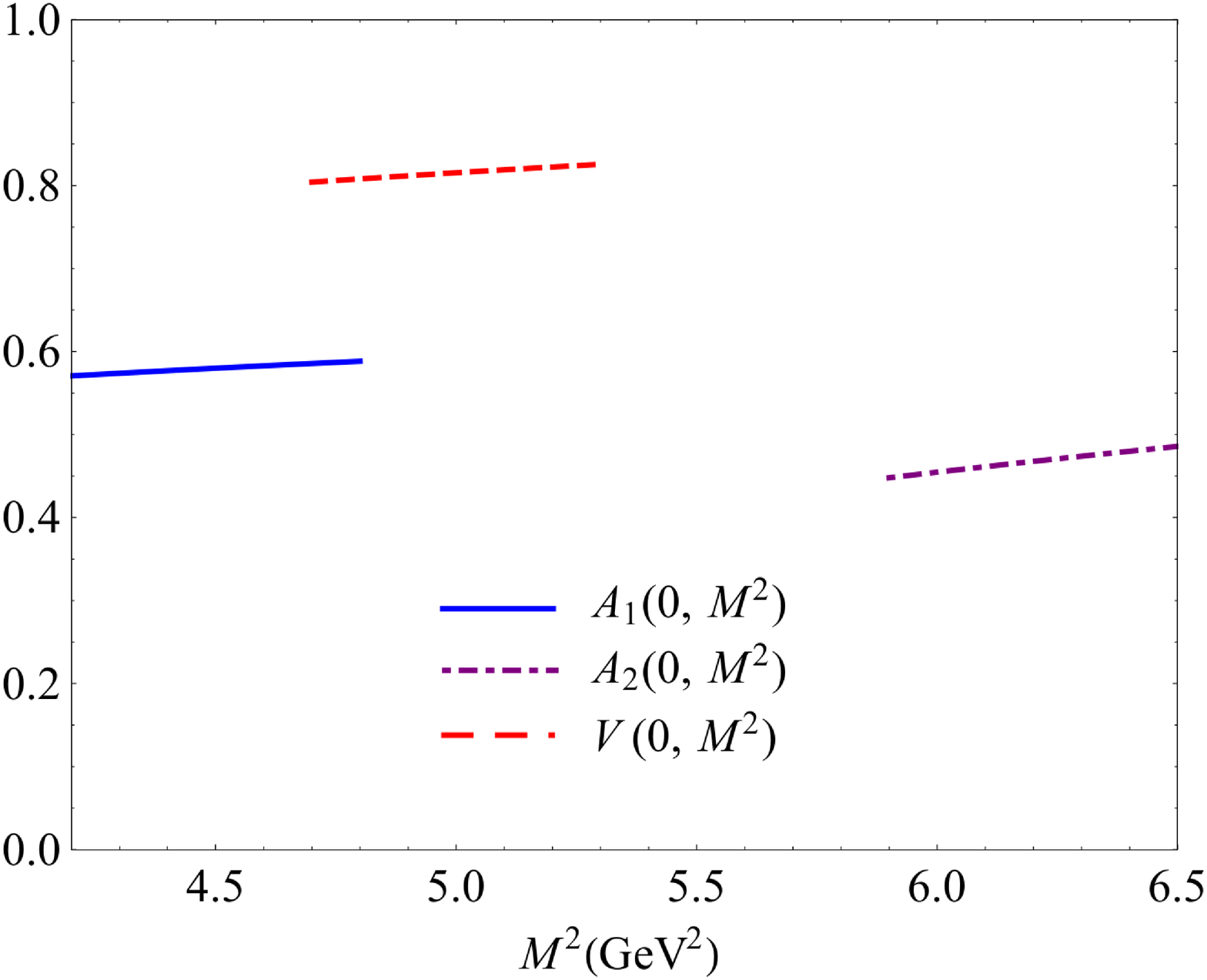}
\caption{The determined Borel windows for the TFFs at the large recoil point, $A_{1,2}(0)$ and $V(0)$. }
\label{Fig:FM2}
\end{figure}

Following those criteria, the determined Borel windows are $M^2= 4.5(3) {\rm GeV}^2$ for $A_1$, $M^2 = 6.2(3) {\rm GeV}^2$ for $A_2$, and $M^2 = 5.0(3) {\rm GeV}^2$ for $V$, respectively. More explicitly, we present the Borel windows for those TFFs at the large recoil point $q^2=0$ in Fig.~\ref{Fig:FM2}, which shows that the TFFs change slightly within the Borel windows, being consistent with conventionally adopted qualitative criterion that the TFF should be flat within the Borel window (since the physical observable should not depend on this artificial parameter). Contributions for different LCDAs for those TFFs are presented in Table~\ref{Tab:TFF01}. It shows that the $\delta^0$-order twist-2 LCDA $\phi_{2;\rho}^\bot$ provides dominate contribution, while the contributions from the $\delta^2$-order high-twist LCDAs are small.

\begin{table}[htb]
\centering
\caption{Contributions from the LCDAs with various twist structures for the $D\to\rho$ TFFs $A_{1,2}(0)$ and $V(0)$, in which only the main non-zero contributions are listed.}
\label{Tab:TFF01}
\begin{tabular}{ c  c  c  c  }
\hline
&~~~~~~$A_1(0)$~~~ &~~~ $A_2(0)$~~~ & ~~~$V(0)$~~~ \\
\hline
$\phi_{2;\rho}^\bot$ & 0.539    & 1.067    & 0.989    \\
$\psi_{3;\rho}^\|  $ & 0.089    & $-0.187$ & /        \\
$\phi_{4;\rho}^\bot$ & $-0.011$ & $-0.157$ & $-0.174$ \\
$I_L$                & $-0.016$ & $-0.091$ & /        \\
$H_3$                & $-0.021$ & $-0.170$ & /        \\
$\Phi_{3;\rho}^\bot$ & 	/       & $0.0003$ & /        \\
Total                & 0.580    & 0.468    & 0.815    \\
 \hline
\end{tabular}
\end{table}

\begin{table}[b]
\centering
\caption{The $D\to \rho$ TFFs $A_{1,2}(q^2)$ and $V(q^2)$ at the large recoil region $q^2\simeq 0$. The errors are squared averages of all the mentioned error sources. As a comparison, we also present the prediction from various methods.}
\label{Tab:Tform factor0}
\begin{tabular}{ c  c  c  c  }
\hline
& $A_1(0)$ & $A_2(0)$ & $V(0)$ \\
\hline
This work & $0.580^{+0.065}_{-0.050}$ &	$0.468^{+0.052}_{-0.053}$ & $0.815^{+0.070}_{-0.051}$  \\
CLEO2013 \cite{CLEO:2011ab} & $0.56(1)^{+0.02}_{-0.03}$ &	$0.47(6)(4)$ & $0.84(9)^{+0.05}_{-0.06}$  \\
3PSR~\cite{Ball:1993tp}	& $0.5(2)$	& $0.4(1)$ & 	$1.0(2)$ \\
HQETF-I~\cite{Wang:2002zba} & $0.57(8)$ & $0.52(7)$ & $0.72(10)$\\
HQETF-II~\cite{Wu:2006rd} & $0.599^{+0.035}_{-0.030}$ & $0.372^{+0.026}_{-0.031}$ & $0.801^{+0.044}_{-0.036}$\\
RHOPM \cite{Wirbel:1985ji}& 	0.78 & 	0.92	& 1.23 \\
QM-I \cite{Isgur:1988gb}& 	0.59& 	0.23	& 1.34  \\
QM-II~\cite{Melikhov:2000yu} & 0.59 & 0.49 & 0.90 \\
LFQM~\cite{Verma:2011yw}       &  $0.60(1)$   &  $0.47(0)$  &  $0.88(3)$ \\
HM$\chi$T~\cite{Fajfer:2005ug} & 0.61 & 0.31 & 1.05 \\
Lattice \cite{Lubicz:1991bi} & 	$0.45(4)$ & $0.02(26)$ & $0.78(12)$  \\
Lattice \cite{Bernard:1991bz}& 	$0.65(15){^{+0.24}_{-0.23}}$ & $0.59(31){^{+0.28}_{-0.25}}$ & 	 $1.07(49)(35)$ \\
 \hline
\end{tabular}
\end{table}

Table~\ref{Tab:Tform factor0} shows the $D\to\rho$ TFFs at the large recoil region $q^2 \rightsquigarrow 0~{\rm GeV^2}$, where the uncertainties are squared averages of all the mentioned error sources for the LCSR approach. As a comparison, we also present the predictions from various approaches in Table~\ref{Tab:Tform factor0}, i.e. from the CLEO collaboration~\cite{CLEO:2011ab}, the 3PSR~\cite{Ball:1993tp}, the HQEFT~\cite{Wang:2002zba, Wu:2006rd}, RHOPM~\cite{Wirbel:1985ji}, the QM~\cite{Isgur:1988gb, Melikhov:2000yu}, the LFQM~\cite{Verma:2011yw}, the HM$\chi$T~\cite{Fajfer:2005ug}, and the Lattice QCD predictions~\cite{Lubicz:1991bi, Bernard:1991bz}, respectively.

\begin{table}[t]
\centering
\caption{Resonance masses of quantum number $J^P$ as indicated necessary for the parameterisation of $D\to\rho$ TFFs $A_{1,2}$ and $V$.}
\label{Tab:mRi}
\begin{tabular}{c c c}
\hline
~~~ $F_i$~~~ &~~~ $J^P$~~~ &~~~ $m_{R,i}/{\rm GeV}$~~~  \\
\hline
%$A_0$     & $0^-$    & 1.870       \\
$V  $     & $1^-$    & 2.007      \\
$A_1,A_2$ & $1^+$    & 2.427      \\
\hline
\end{tabular}
\end{table}

\begin{figure*}[htb]
\centering
\includegraphics[width=0.45\textwidth]{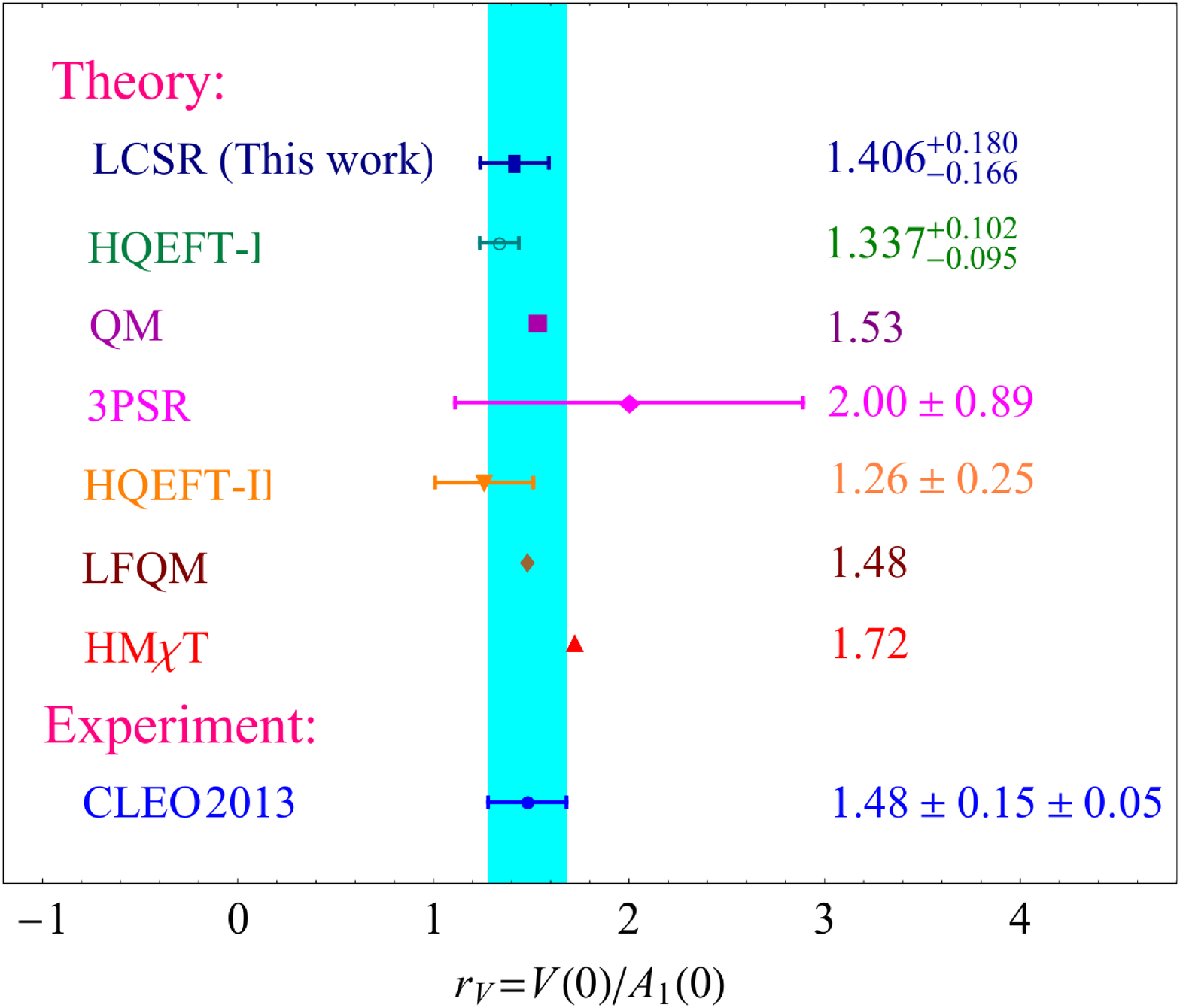}~\includegraphics[width=0.45\textwidth]{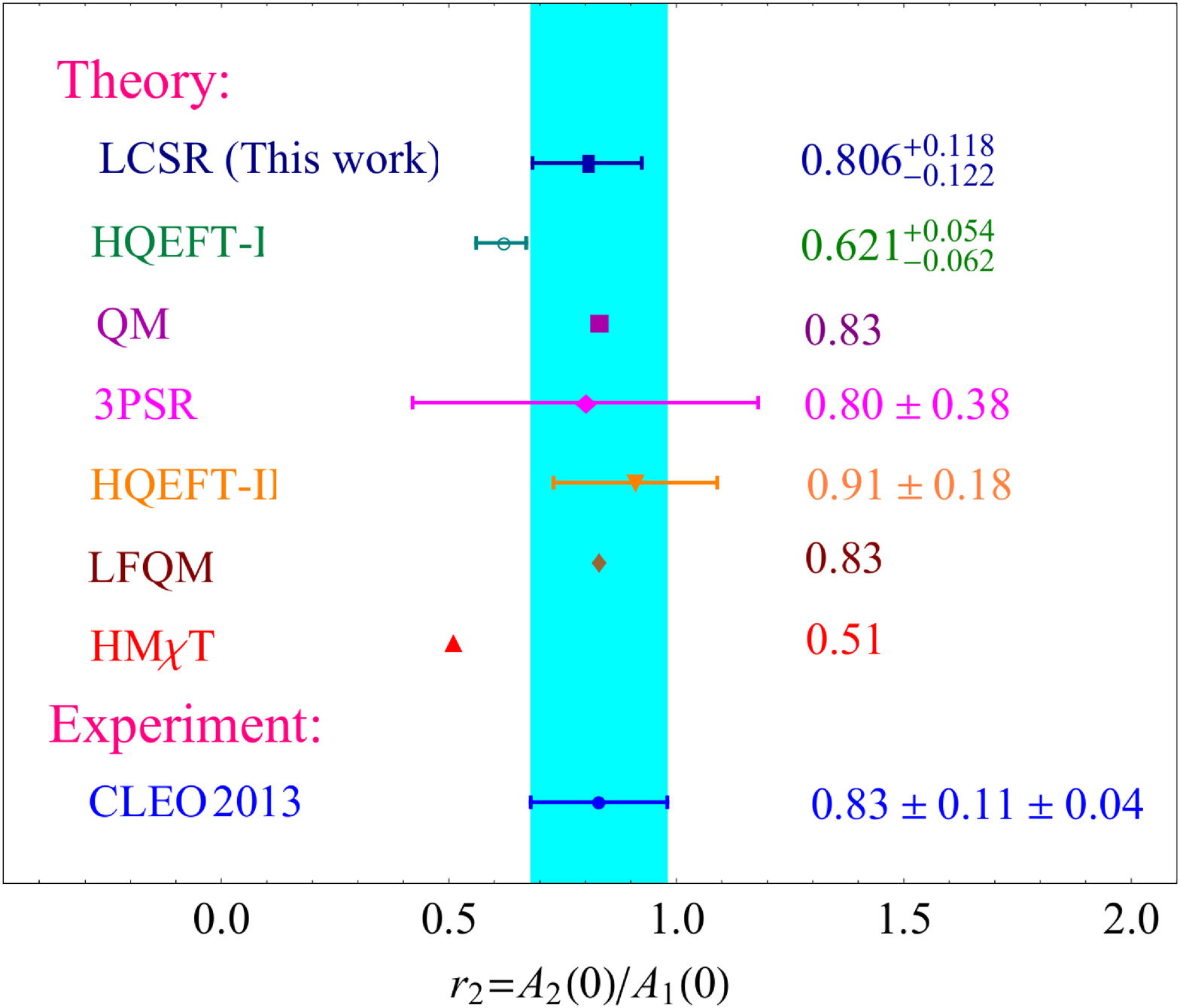}
\caption{Comparison of theoretical predictions for the ratios $r_2$ and $r_V$ within various approaches. The CLEO measurements are presented as a comparison. } \label{Fig_Tform factorratio}
\end{figure*}

Table~\ref{Tab:Tform factor0} shows that the TFFs under many approaches are consistent with each other within reasonable errors. To show the relative importance of various TFFs within different approaches more clearly, we present a comparison of the ratios $r_2$ and $r_V$ in Fig.~\ref{Fig_Tform factorratio}. The LCSR uncertainties for the TFFs are $\left(^{+13\%}_{-12\%}\right)$ for $r_V$ and $\left(^{+15\%}_{-15\%}\right)$ for $r_2$, which are much smaller than the previous 3PSR predictions (which are $\pm45\%$ and $\pm 48\%$~\cite{Ball:1993tp}, respectively). Thus, by using the LCSR approach, more accurate QCD SR predictions can be obtained.

\begin{figure*}[htb]
\centering
\includegraphics[width=0.33\textwidth]{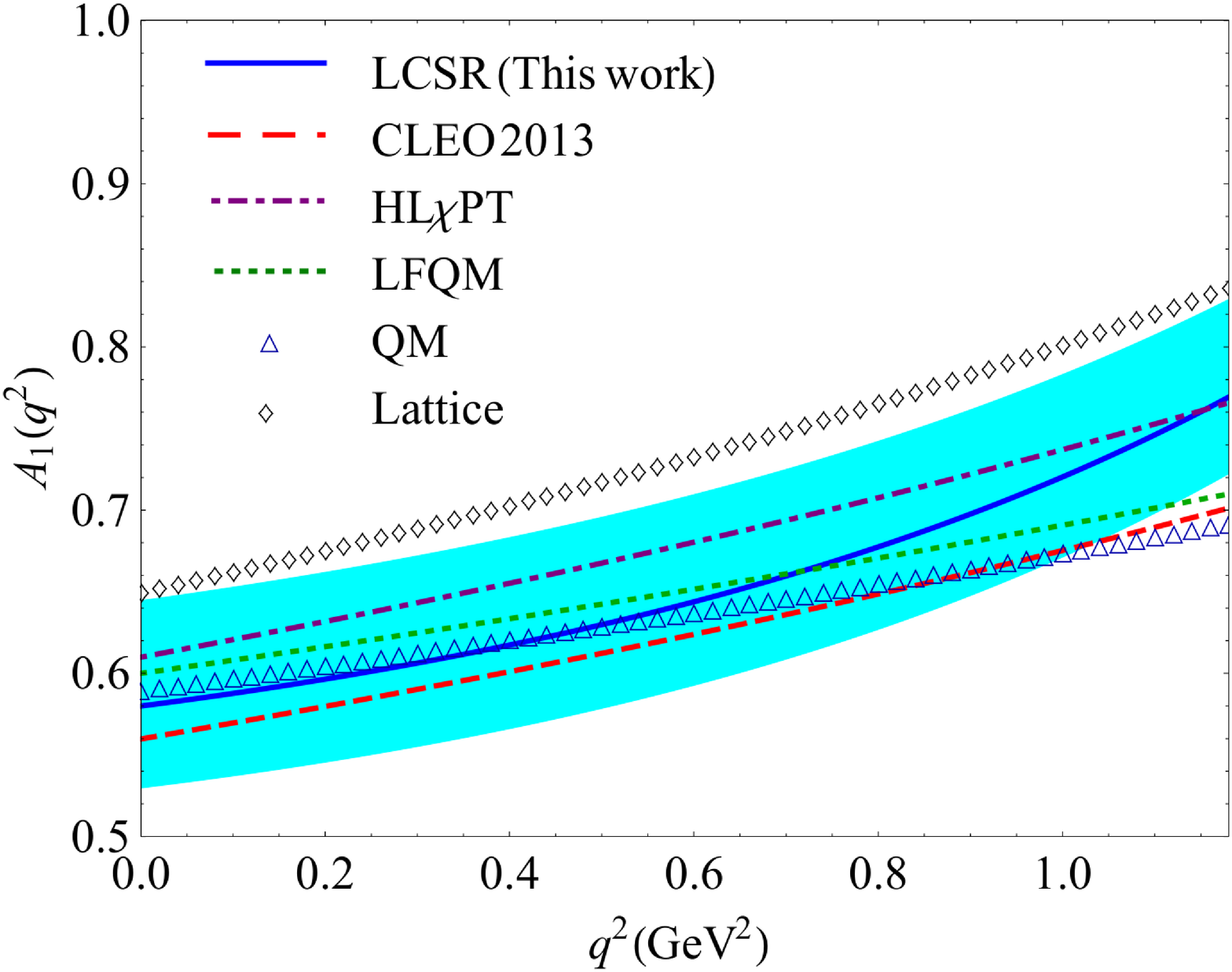}~\includegraphics[width=0.33\textwidth]{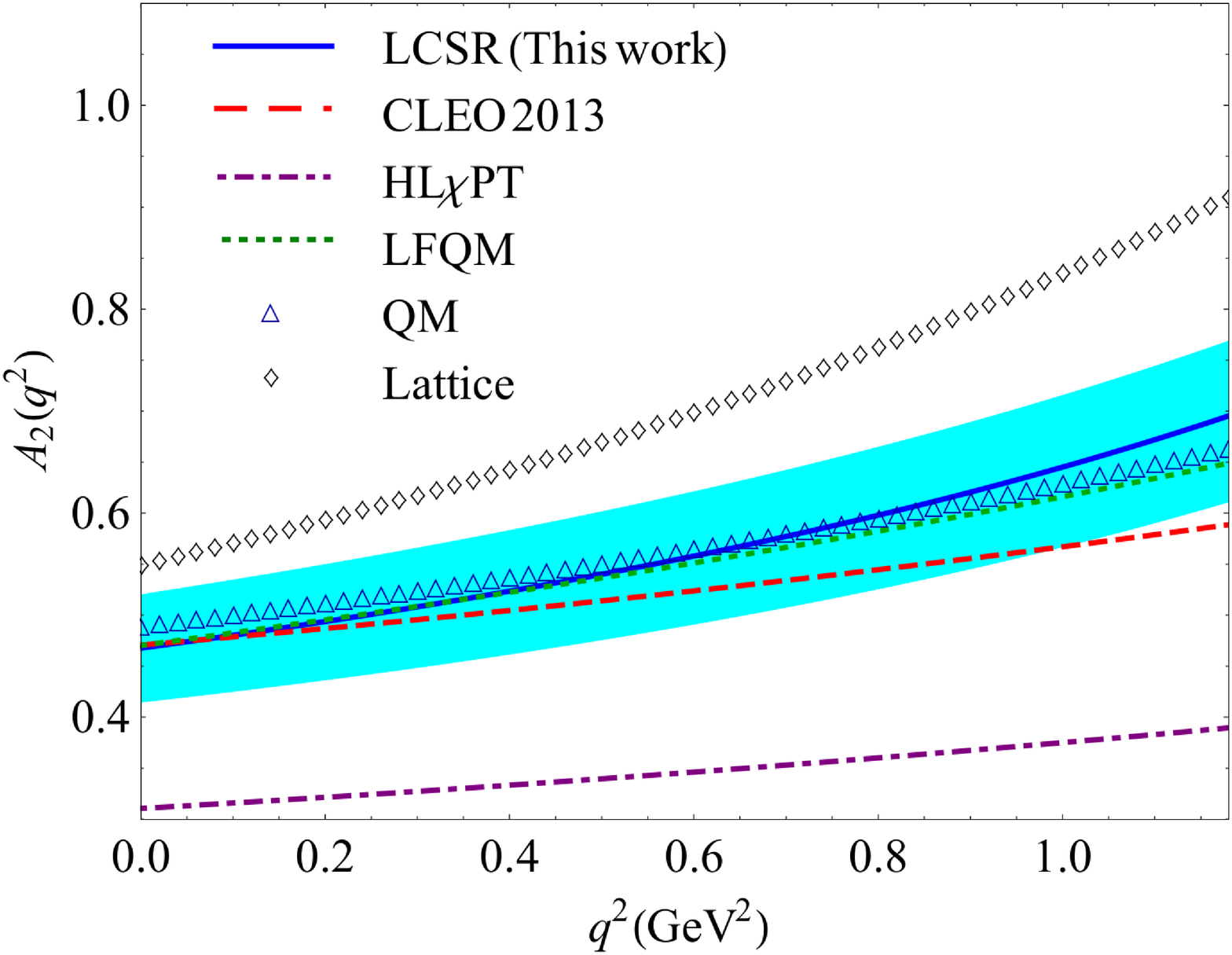}~\includegraphics[width=0.33\textwidth]{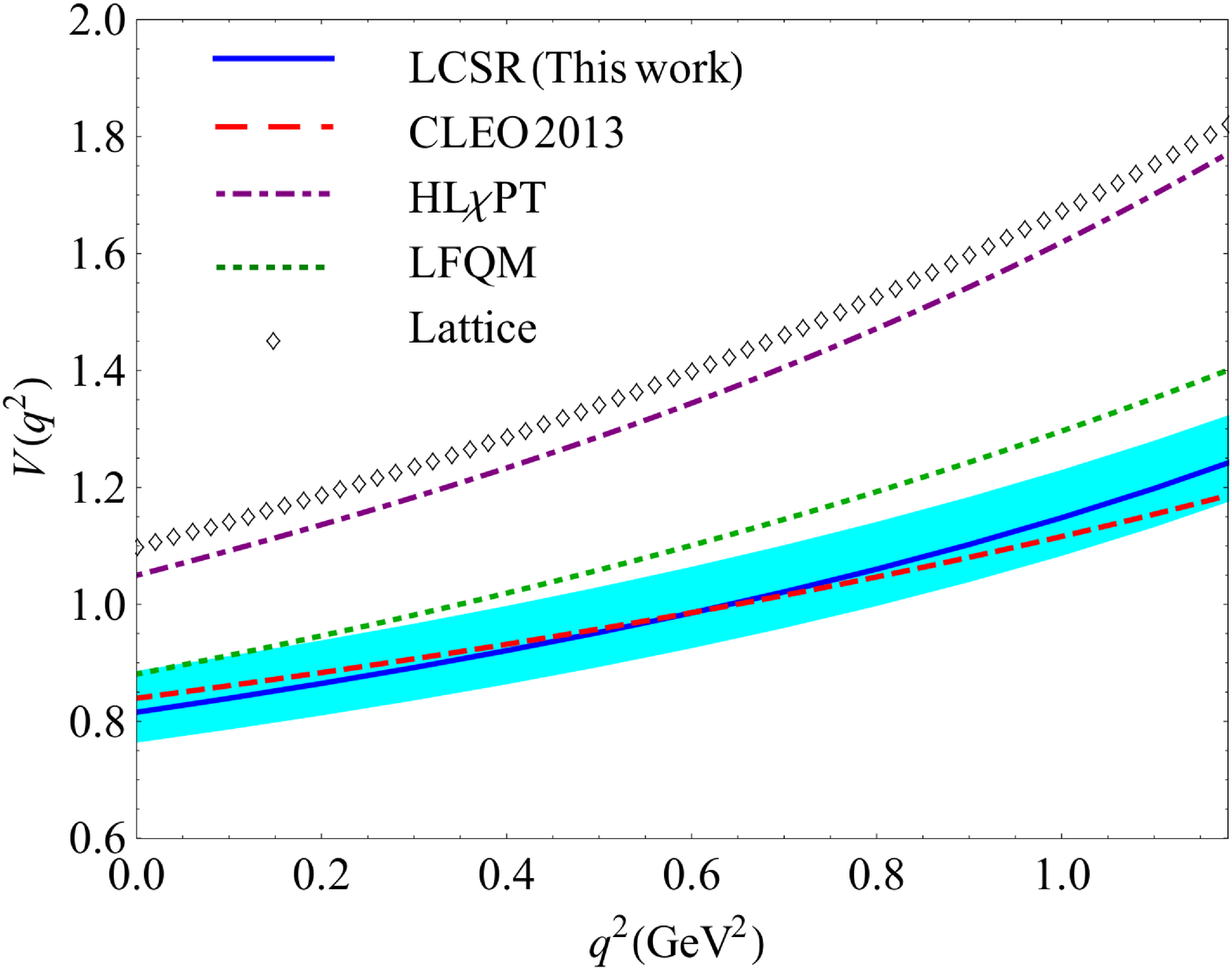}
\caption{The extrapolated $D\to\rho$ TFFs $A_{1,2}(q^2)$ and $V(q^2)$, in which the shaded bands are squared average of those from the mentioned error sources. As a comparison, we also present the central value of the QM~\cite{Melikhov:2000yu}, the LFQM~\cite{Verma:2011yw}, the HL$\chi$PT~\cite{Fajfer:2005ug}, the Lattice QCD predictions~\cite{Flynn:1997ca}, and CLEO measurments~\cite{CLEO:2011ab} in those figures. }
\label{Fig:Tform factor}
\end{figure*}

The physically allowable range for the TFFs is $0\leq q^2 \leq (m_D-m_\rho)^2=1.18 {\rm GeV}^2$. Theoretically, the LCSRs for the $D\to \rho$ TFFs are applicable in low and intermediate $q^2$-regions, e.g. $q^2\in[0,0.8]\;{\rm GeV}^2$. We can extrapolate them to whole $q^2$-regions via a rapidly converging series over the $z(t)$-expansion~\cite{Khodjamirian:2010vf, Straub:2015ica}
\begin{eqnarray}
F_i(q^2) = P_i (q^2) \sum_{k=0,1,2}a_k^i [z(q^2)-z(0)]^k,
\end{eqnarray}
where
\begin{eqnarray}
z(t)=\frac{\sqrt{t_+ - t}-\sqrt{t_+ - t_0}}{\sqrt{t_+ - t}+\sqrt{t_+ - t_0}}
\end{eqnarray}
with $t_\pm=(m_D\pm m_\rho)^2$, $t_0=t_+(1-\sqrt{1-t_-/t_+})$, and the $F_i$ are three TFFs $A_{1,2}$ and $V$, respectively. The $P_i (q^2) = (1-q^2/m_{R,i}^2)^{-1}$ is a simple pole corresponding to the first resonance in the spectrum. The appropriate resonance masses are given in Table~\ref{Tab:mRi}. The parameters $a_k^i$ can be fixed by requiring $\Delta < 0.1\%$, and the results are put in Table~\ref{analytic}. Here the parameter $\Delta$ is introduced to measure the quality of extrapolation,
\begin{equation}
\Delta=\frac{\sum_t\left|F_i(t)-F_i^{\rm fit}(t)\right|} {\sum_t\left|F_i(t)\right|}\times 100, \label{delta}
\end{equation}
where $t\in[0,\frac{1}{40},\cdots,\frac{40}{40}]\times 0.8 {\rm GeV}^2$.

\begin{table}[htb]
\caption{The fitted parameters $a^i_{1,2}$ for the $D\to \rho$ TFFs $F_i$, in which all the LCSR parameters are set to be their central values. $\Delta$ is the measure of the quality of extrapolation.} \label{analytic}
\begin{tabular}{ c  c c c }
\hline
          & ~~~~$A_1$~~~~ & ~~~~$A_2$~~~~ & ~~~~$V$~~~~ \\ \hline
$a_1^i$   & 0.711 & $-1.149$ & $-0.797$ \\
$a_2^i$   & 29.23 & $15.108$ & $10.370$ \\
$\Delta$  &  0.05\% &   0.04\%   & 0.01\%     \\
\hline
\end{tabular}
\end{table}

The extrapolated TFFs in whole $q^2$-region are presented in Fig.~\ref{Fig:Tform factor}, where the shaded bands are uncertainties from various input parameters. As a comparison, we also give the results from various approaches, which are from the CLEO Collaboration~\cite{CLEO:2011ab}, the QM~\cite{Melikhov:2000yu}, the LFQM~\cite{Verma:2011yw}, the HL$\chi$PT~\cite{Fajfer:2005ug}, and the Lattice predictions~\cite{Flynn:1997ca}, respectively. The CLEO Collaboration only issues the TFFs at large recoil region, and the present CLEO curves are fitted ones from the large energy chiral-quark model~\cite{Palmer:2013yia}.

\subsection{The semi-leptonic decay $D\to\rho e\nu_e$}

\begin{table}[htb]
\begin{center}
\caption{Total decay width $1/|V_{\rm cd}|^2 \times \Gamma$, the ratio of longitudinal and transverse decay width  $\Gamma^L/ \Gamma^T$, and the ratio of positive and negative decay width $\Gamma^+/ \Gamma^-$.}
\label{Tab:GammaRatio}
\begin{tabular}{c  c  c c }
\hline
& $1/|V_{\rm cd}|^2 \times \Gamma$ & $\Gamma^{\rm L}/ \Gamma^{\rm T}$ & $\Gamma^+/ \Gamma^-$ \\
\hline
This paper	               & $55.45^{+13.34}_{-9.41}$ & $1.18^{+0.14}_{-0.13}$ & $0.22^{+0.04}_{-0.03}$       \\
3PSR~\cite{Ball:1993tp}      & $15.80\pm4.61$           & $1.31\pm0.11$          & $0.24\pm0.03$ \\
HQEFT~\cite{Wang:2002zba} & $71\pm14$           & $1.17\pm0.09$          & $0.29\pm0.13$ \\
RHOPM\cite{Wirbel:1985ji}  & $90.83$                  & 0.91                   & 0.19    \\
QM~\cite{Isgur:1988gb}        & $88.86$                  & 1.33                   & 0.11    \\
Lattice\cite{Lubicz:1991bi}       & $54.63\pm12.51$          & $1.86\pm0.56$          & 0.16    \\
Lattice\cite{Bernard:1991bz}      & 71.75                    & 1.10                   & 0.18    \\
\hline
\end{tabular}
\end{center}
\end{table}

By using the extrapolated $D\to\rho$ TFFs, we calculate the total decay width $1/|V_{\rm cd}|^2 \times \Gamma$, the ratio of longitudinal and transverse decay width  $\Gamma^{\rm L}/ \Gamma^{\rm T}$ for the $D\to\rho$ semileptonic decays, and the ratio of positive and negative decay width $\Gamma^+/ \Gamma^-$. The results are presented in Table~\ref{Tab:GammaRatio}, where the results under various approaches have also been presented. Table~\ref{Tab:GammaRatio} shows that our LCSR predictions for $1/|V_{\rm cd}|^2 \times \Gamma$, $\Gamma^{\rm L}/ \Gamma^{\rm T}$ and $\Gamma^+/ \Gamma^-$ are consistent with other approaches within errors, only the value of $1/|V_{\rm cd}|^2 \times \Gamma$ is quite larger than the 3PSR prediction~\cite{Ball:1993tp}.

\begin{table}[htb]
\begin{center}
\caption{The branching ratios of the semileptonic decays $D^0\to \rho^- e^+ \nu_e$ and $D^+ \to \rho^0 e^+ \nu_e$ (in unit: $10^{-3}$). As a comparison, we also present the results of CLEO Collaboration~\cite{CLEO:2011ab, Huang:2005iv}, 3PSR~\cite{Ball:1993tp}, HQEFT~\cite{Wang:2002zba}, NWA~\cite{Shi:2017pgh} with HQEFT~\cite{Wu:2006rd} and LFQM~\cite{Verma:2011yw}, FK~\cite{Fajfer:2005ug} and ISGW2~\cite{Scora:1995ty}. }
\label{Table:Branching_se}
\begin{tabular}{c c c }
\hline
Decay Mode	& $D^0\to \rho^- e^+ \nu_e$          & $D^+ \to \rho^0 e^+ \nu_e$         \\ \hline
This paper	& $1.749^{+0.421}_{-0.297}\pm 0.006$ & $2.217^{+0.534}_{-0.376}\pm 0.015$ \\
CLEO2005~\cite{Huang:2005iv} & $1.94\pm0.39\pm0.13$ & $2.1\pm0.4\pm0.1$ \\
CLEO2013~\cite{CLEO:2011ab}     & $1.77\pm0.12\pm0.10$      	     & $2.17\pm0.12^{+0.12}_{-0.22}$      \\
3PSR \cite{Ball:1993tp}	& $0.5\pm0.1$	& --    \\
HQEFT~\cite{Wang:2002zba} & $1.4\pm0.3$ & -- \\
NWA~\cite{Shi:2017pgh}+HQEFT~\cite{Wu:2006rd}  & $1.67\pm0.27$ & $2.16\pm0.36$ \\
NWA~\cite{Shi:2017pgh}+LFQM~\cite{Verma:2011yw}   & $1.73\pm0.07$  & $2.24\pm0.09$ \\
FK \cite{Fajfer:2005ug} & 2.0	& 2.5 \\
ISGW2 \cite{Scora:1995ty} & 1.0 & 1.3  \\
\hline
\end{tabular}
\end{center}
\end{table}

As a further step, we calculate the branching ratios for the two $D\to\rho$ semileptonic decays. One is the $D^0$-type decay via the process $D^0 \to \rho^- e^+\nu_e$ with the lifetime $\tau(D^0)= 0.410\pm 0.002$ ps, and the other one is the $D^+$-type decay via the process $D^+ \to \rho^0 e^+\nu_e$ with the lifetime $\tau(D^+)= 1.040\pm 0.007$ ps~\cite{Tanabashi:2018oca}. The results are given in Table~\ref{Table:Branching_se}, where the first uncertainty is squared average of the mentioned error sources, and the second uncertainty is from the experimental errors for the lifetime. As a comparison, we also list the branching ratios derived from various approaches in Table~\ref{Table:Branching_se}. It indicates that a smaller $1/|V_{\rm cd}|^2 \times \Gamma$ predicted by 3PSR leads to a smaller branching ratio. This explains why the previous SR prediction is inconsistent with other approaches. However, by using the LCSR approach, we observe that a more reasonable and accurate SR prediction can be achieved. The LCSR predictions for the branching ratios of the two $D\to\rho$ semileptonic decays also show better agreement with the CLEO measurements.

\subsection{The radiative decay $D\to\rho \gamma$}

\begin{table}[htb]
\begin{center}
\caption{The branching ratios for $D\to \rho\gamma$ decays (in unit: $10^{-5}$). As a comparison, we also present other theoretical predictions.}
\label{Table:Br_rediative deccay}
\begin{tabular}{c c c }\hline
	                              & ~${\mathcal B}(D^0\to \rho^0 \gamma)$&${\mathcal B}(D^+ \to \rho^+ \gamma)$~~~\\ \hline
This paper~(SM)	                  & $1.744^{+0.598}_{-0.704}$   & $5.034^{+0.939}_{-0.958}$ \\
QCD SM~\cite{Khodjamirian:1995uc} & $0.38$                      & $0.46$                    \\
Hybrid~\cite{deBoer:2017que}      & $0.606\pm0.565$             & $1.174\pm1.157$           \\
FS~\cite{Fajfer:1997bh}           & $0.55\pm0.45$               & $3.35\pm2.95$             \\
Pole Diagrams and VMD~\cite{Burdman:1995te}& $0.3\pm0.2$        & $4\pm2$                   \\
Belle Collaboration~\cite{Abdesselam:2016yvr}    & $1.77\pm0.31$            &  $-$                      \\ \hline
\end{tabular}
\end{center}
\end{table}

After inputting the $D\to \rho$ TFFs into the parity conserving and parity violating effective couplings ${\cal A}_{\rm PV, PC}^\rho$, we get the branching ratio for the two $D$ meson radiative processes $D^0\to \rho^0\gamma$ and $D^+\to\rho^+\gamma$. The results are given in Table~\ref{Table:Br_rediative deccay}, where the uncertainties are squared average of the theoretical and experimental error sources. As a comparison, we also listed the branching ratios derived from various approaches. It indicates that the LCSR predictions for the branching ratios for radiative decay $D^0\to \rho^0 \gamma$ shows a better agreement with the Belle Collaboration result, which is larger than other theoretical predictions.

For the direct CP asymmetry of the $D\to \rho\gamma$ decay, we recall that the maximal value of Eq.\eqref{Eq:CP_asymmetry} can be reached in the limit of maximal constructive interference (namely of $\pm \pi/2$ strong phase difference) of the amplitudes with different weak phases. This way we get the upper limit of our predictions for the direct CP asymmetry.

The Wilson coefficient $C_7^{(\prime)\rm eff}$ mainly comes from the hard spectator interaction and weak annihilation contributions within the framework of SM, which results in $C_7^{\rm SM}(m_c) \in [-0.00949+0.0014 i, -0.00019+0.002 i]$~\cite{deBoer:2017que}. By taking input parameters into Eq.\eqref{Eq:CP_asymmetry}, we obtain the SM prediction for the CP asymmetry of $D^0\to \rho^0\gamma$, e.g.
\begin{eqnarray}
A_{CP}^{\rm SM} &=& \big[1.329 (\pm0.234)_{C_7} (\pm0.089)_{m_c} \left(^{+0.094}_{-0.073}\right)_{\rm F.F.}
\nonumber\\
&& \left(^{+0.295}_{-0.204}\right)_{a_2}  \left(^{+0.052}_{-0.049}\right)_{\rm D.C.} \big]\times 10^{-2}. \\
&=& \left(1.329^{+0.402}_{-0.335}\right)\times 10^{-2}.  \label{Eq:Value_ACP}
\end{eqnarray}
In the first line, the separate uncertainties are caused by the errors of the quantities $C_7$, $m_c$, TFFs (F.F.), $a_2$ and decay constant (D.C.), respectively, and in the second line, all the errors are added up in quadrature. The central value of the SM prediction is smaller than the Belle result, $A^{\rm exp}_{CP} = 0.056\pm 0.152\pm0.006$~\cite{Abdesselam:2016yvr} by almost $3$ times. Because of the large statistical error for the present Belle measurements, the SM prediction roughly agrees with the Belle result within errors.

One may hope that the possible discrepancy can be accommodated by a well-motivated extension of the SM. To quantify the size of the Wilson coefficients, one can normalize the effective Hamiltonian within new-physics contributions as
\begin{eqnarray}
{\cal H}^{\rm eff-NP} = \frac{G_F}{\sqrt 2} \sum_i C_i {\cal Q}_i + h.c.,~\label{EQ:H}
\end{eqnarray}
where the complete list of potentially relevant operators can be found in Ref.~\cite{deBoer:2015boa}. The Wilson coefficients $\delta C_{7,8}^{(\prime)}(M)$ are generically related beyond SM models, with $M$ denotes the matching scale. The initial conditions of the four operators are assuming $M > m_t$, taking into account the renormalization group evolution of the operators at the leading log level, leads to
\begin{eqnarray}
&&C^{(\prime)}_7 (m_c) = \tilde\eta[\eta C^{(\prime)}_7 (M) + 8(\eta -1)C^{(\prime)}_8(M)],\label{Eq:C7} \\
&&C^{(\prime)}_8 (m_c) = \tilde\eta C^{(\prime)}_8 (M), \label{Eq:C8}
\end{eqnarray}
where the coefficients $\eta$ and $\tilde\eta$ can be found in Ref.\cite{Buras:1999da}. A non-vanishing value for $\Delta A_{\rm CP} =  A_{\rm CP}(K^+ K^-) - A_{\rm CP}(\pi^+ \pi^-)$ has been observed by LHCb and CDF collaborations~\cite{Aaij:2011in, Chatrchyan:2011fq}, which could be used to restrict the new physics contribution from the operator ${\cal Q}_8$ by using the relationship $\Delta A_{\rm CP} \approx -1.8 |{\rm Im} [C_8^{\rm NP}(m_c)]$~\cite{Giudice:2012qq}. Considering the world average value $\Delta A_{\rm CP}^{\rm exp} = -(0.67\pm0.16)\%$~\cite{Aaltonen:2011se}, one can get $|{\rm Im} [C_8^{\rm NP}(m_c)]| \approx 0.4\times 10^{-2}$. Furthermore the renormalization group evolution implies $|{\rm Im}[C_7^{\rm NP}(m_c)]| \approx |{\rm Im}[C_8^{\rm NP}(m_c)]|$ if the initial scale $M$ is set to be around $1$ TeV, which, for instance, could be happened for super-symmetry theory.
Taking into account the uncertainties in determining $|{\rm Im}[C_8^{\rm NP}(m_c)]|$ and the uncertainties from the initial conditions of $|C_7^{\rm NP}(M)|$, one can obtain a conservative range $(0.2-0.8) \times 10^{-2}$ for $|{\rm Im}[C_7^{\rm NP}(m_c)]|$~\cite{Isidori:2012yx}. Thus we obtain a prediction for the CP asymmetry of $D^0\to \rho^0\gamma$,
\begin{eqnarray}
A_{CP}^{\rm NP} &=& \big[3.907 (\pm 2.344)_{C_7} (\pm 0.260)_{m_c} \left(^{+0.276}_{-0.215}\right)_{\rm F.F.}
\nonumber\\
&& \left(^{+0.868}_{-0.601}\right)_{a_2} \left(^{+0.153}_{-0.145}\right)_{\rm D.C.} \big]\times 10^{-2} \\
&=& \left(3.907^{+2.533}_{-2.448}\right) \times 10^{-2}.
\end{eqnarray}
In the second line, all the errors are added up in quadrature. This value agrees with the Belle Collaboration result within errors.

\section{Summary}\label{section:4}

In the paper, we have investigated the $D \to \rho$ TFFs within the LCSR approach. As shown by Table \ref{Tab:Tform factor0} and Fig.\ref{Fig_Tform factorratio}, more accurate QCD SR predictions for the TFFs $A_{1,2}(q^2)$ and $V(q^2)$ can be achieved by applying the LCSR approach other than the 3PSR approach. To compare with the CLEO measurements, the LCSR approach can give reasonable explanations for the $D \to \rho$ TFFs. The pQCD factorization approach is applicable in large recoil region $q^2 \rightsquigarrow 0$ and the lattice QCD approach is applicable in very large $q^2$-region, thus the extrapolation of the results under those two approaches shall be strongly model dependent. The LCSR prediction provides a bridge between the pQCD and lattice QCD predictions, since it is applicable for a wider range, i.e. in both small and intermediate $q^2$-region.

After extrapolating the $D\to\rho$ TFFs to whole $q^2$-region, the LCSR predictions for the branching ratios of the two $D$ meson semileptonic decays are ${\cal B} (D^0\to \rho^- e^+ \nu_e) = (1.749^{+0.421}_{-0.297}\pm 0.006)\times 10^{-3}$ and ${\cal B} (D^+ \to \rho^0 e^+ \nu_e) = (2.217^{+0.534}_{-0.376}\pm 0.015)\times 10^{-3}$, respectively, which agree with the CLEO measurements within errors. And as shown by Table \ref{Table:Br_rediative deccay}, the branching ratios of the $D$ meson radiative decay $D^0 \to \rho^0 \gamma$ and $D^+ \to \rho^+ \gamma$ also show good agreement with the experimental results in comparing with other theoretical predictions.

Based on the short and long distance parity conserving and parity violating amplitudes, i.e. $( \mathcal{A}_{\mathrm{PV/PC}}^{\rho})^{\rm s.d./l.d.}$, and in the limit of maximal constructive interference of the amplitudes, we get the SM prediction for the CP asymmetry of $D^0\to \rho^0\gamma$, $ A_{CP}^{\rm SM}=(1.329^{+0.402}_{-0.335})\times 10^{-2}$, which has large discrepancy with the Belle Collaboration data. By taking $|{\rm Im}[C_7^{\rm NP}(m_c)]| = (0.2-0.8) \times 10^{-2}$ for the NP Wilson coefficient, we obtain the NP prediction, $A_{CP}^{\rm NP} = (3.907^{+2.533}_{-2.448})\times 10^{-2}$, which agrees with the Belle Collaboration result within errors. Since the short-distance amplitude is relatively much smaller than the long-distance amplitude, the NP effect induced by the operator ${\cal Q}_8$ shall give negligible effect (less than 0.1\%) to the SM predictions of the branching ratios. Thus we think that the LCSR approach can provide a self-consistent way to deal with the $D$ meson decays, and the $D$ meson involved processes could be adopted for testing NP beyond the SM.

\hspace{1cm}

{\bf Acknowledgments}: This work was supported in part by the National Science Foundation of China under Grant No.11881240255, No.11765007 and No.11625520, No.11847301, the Project of Guizhou Provincial Department of Science and Technology under Grant No.KY[2017]1089 and No.KY[2019]1171, the China Postdoctoral Science Foundation under Grant No.2019TQ0329, the Project of Guizhou Minzu University under Grant No.GZMU[2019]YB19.

\end{document}